\documentclass[doublecol]{epl2} 

\providecommand{\hm}{\pmb\bm}
\providecommand{\mbf}{\mathbf}

\usepackage[latin1]{inputenc}
\usepackage{mathtools}
\usepackage{amssymb}
\usepackage{bm}

\RequirePackage{xspace}

\renewcommand*{\d}[1]{\mathrm{d}#1}

\newcommand*{\dalembert}{\square}

\newcommand*{\im}{{\mathrm{i}}}

\renewcommand*{\vec}[1]{\mbf{#1}}

\newcommand*{\grad}{\hm{\nabla}}
\newcommand*{\bdot}{\bm\cdot}
\renewcommand*{\div}{\grad\bdot}
\newcommand*{\cross}{\bm\times}
\newcommand*{\curl}{\grad\cross}

\title{Photon orbital angular momentum and mass in a plasma\\
 vortex}
\shorttitle{POAM and mass in a plasma vortex} 

\author{F.\,Tamburini\inst{1} \and A.\,Sponselli\inst{1} \and B.\,Thid\'e\inst{2} \and J.\,T. Mendon\c{c}a\inst{3}}
\shortauthor{F. Tamburini \etal}

\institute{                    
  \inst{1} Department of Astronomy,
 University of Padova,
 vicolo dell' Osservatorio~3,
 I-33122 Padova,
 Italy\\
  \inst{2} Swedish Institute of Space Physics,
 Physics in Space,
 {\AA}ngstr\"{o}m Laboratory,
 P.\,O.~Box~537,
 SE-751\,21,
 Uppsala,
 Sweden\\
 \inst{3}
  IPFN and CFIF, Instituto Superior T\'{e}cnico,
 Av.~Rovisco Pais 1,
 1049-001 Lisboa,
 Portugal
}
\pacs{52.35.We}{Plasma vorticity}

\abstract{We analyse the Anderson-Higgs mechanism of photon mass
acquisition in a plasma and study the contribution to the mass from
the orbital angular momentum acquired by a beam of photons when it
crosses a spatially structured charge distribution. To this end we apply
Proca-Maxwell equations in a static plasma with a particular spatial
distribution of free charges, notably a plasma vortex, that is able to
impose orbital angular momentum (OAM) onto light. In addition to the
mass acquisition of the conventional Anderson-Higgs mechanism, we find
that the photon acquires an additional mass from the OAM and that this
mass reduces the Proca photon mass.}

\begin{document}

\maketitle

\section{Introduction}

Influenced by results derived in 1962 by Schwinger
\cite{PhysRev.125.397}, Anderson showed, in 1963, that a photon
propagating in a plasma acquires a mass, $\mu_\gamma=\hbar\omega_p/c^2$,
where $\omega_p$ is the plasma frequency \cite{PhysRev.130.439}.
In this Anderson-Higgs process the photon acquires an effective
mass because of its interaction with plasmons, or better, a hidden
gauge invariance in the plasma \cite{Mendonca:Book:2001}. In order
to study photons that have acquired an effective mass, it is
convenient to replace Maxwell's equations by Proca-Maxwell equations
\cite{hora1991plasmas,RevModPhys.82.939}. In this Letter, we use this
approach to analyse the contribution to the mass from the orbital
angular momentum acquired by a beam of photons as it traverses a
spatially structured charge distribution.

The orbital angular momentum (OAM) of light is a newly recognised observable
of electromagnetic (EM) fields that is intimately related to
optical vortices (OV's), phase defects embedded in certain
particular light beams. OAM of light beams can be generated by
the imprinting of vorticity onto the phase distribution of the
original beam when it crosses inhomogeneous nonlinear optical
systems \cite{1991PhRvL..67.3749A} or particular spatial structures
such as fork holograms or spiral phase plates. Such beams can be
mathematically described by a superposition of Laguerre-Gaussian (L-G)
modes characterized by the two integer-valued indices $l$ and $p$,
or by Kummer beams \cite{anzolin:033845}. The azimuthal index $l$
describes the number of twists of the helical wavefront and the radial
index $p$ gives the number of radial nodes of the mode.

The EM field amplitude of a generic L-G mode, in a plane perpendicular to
the direction of propagation, is
\begin{align}
\label{eqn:lgmode}
F_{pl}(r,\varphi)
 = \sqrt{\frac{(l+p)!}{4\pi p!}}
 \Big(\frac{r^2}{w^2}\Big)^{\!|l|}\,L_p^{|l|}\Big(\frac{r^2}{w^2}\Big)
 \exp\Big(-\frac{r^2}{2w^2}\Big)e^{\im l\varphi}
\end{align}
obeying the orthogonality condition
\begin{align}
\int_0 ^\infty r\,\d{r} \int_0 ^{2\pi} F_{pl}^* F_{p'l'}\,\d{\varphi} 
 = \delta_{pp'}\delta_{ll'}
\end{align}
where $w$ is the beam waist and $L_n^m$ is the associated Laguerre
polynomial. The phase factor $\exp(-\im l\varphi)$ is associated with
an OAM of $l\hbar$ per photon, and a phase singularity is embedded in
the wavefront, along the propagation axis, with a topological charge
$l$ \cite{1992PhRvA..45.8185A, 2002JOptB...4S..47V}. The intensity
distribution of an L-G mode with $p=0$ has an annular shape with
a central dark hole where the intensity vanishes because of total
destructive interference.

As is well known, not only the linear momentum of light
but also its angular momentum can propagate to infinity
\cite{Jackson:490457,Schwinger&al:Book:1998,Thide:Book:2010}. The
OAM property of the field remains stable during the propagation in
free space and has been experimentally verified down to the single
photon limit \cite{PhysRevLett.88.053601}. It has also been studied
theoretically \cite{2008PhRvA..78e2116T}.

A photon in vacuum has its intrinsic angular momentum, the spin (SAM),
that can assume only two values, $\sigma=\pm1$ corresponding to 
transverse left-hand and right-hand circular polarisation,
respectively. If the photon were massive, then it would carry a third
independent longitudinal polarisation component along the direction
of propagation \cite{Heitler:Book:1954}. The exchange of angular
momentum between a photon beam and a plasma vortex and the possible
excitation of photon angular momentum states in a plasma was analysed
in Ref.~\cite{Mendonca&al:PRL:2008}. There can also be an exchange of
angular momentum between electromagnetic and electrostatic waves in
a plasma due to stimulated Raman and Brillouin scattering processes
\cite{Mendonca&al:PRL:2009}. The properties of plasmons carrying OAM
were analysed in Ref.~\cite{Mendonca&al:POP:2009}.

In this Letter we show that the OAM acquired by a photon in a
spatially structured plasma can be interpreted as an additional
mass-like term that appears in Proca equations. More specifically,
we study the propagation of a photon with wavelength $\lambda$ in a
static helicoidally distributed plasma with step $q_0=\lambda/p$ where
$p$ is an integer \cite{Mendonca&al:PRL:2009}. This apparent mass
term shows that structured spatial and temporal inhomogeneities of
matter distribution can impose properties onto single quanta. This can be
interpreted as a manifestation of a Mach principle in classical
and quantum electrodynamics. Henceforth in this Letter, we use natural units
$e=c=\hbar=G=1$.

\section{Proca equations and OAM in a plasma vortex}

The Maxwell-Proca Lagrangian density $\Lambda$ describes
the kinetic potential of a massive EM field in which there
appears a mass term $\mu_\gamma^2A_\mu A^\mu/2$ because of the
chosen gauge invariance:
\begin{equation}
\Lambda= - \frac{1}{4} F_{\mu\nu}F^{\mu\nu}-j_\mu A^\mu
 + \frac{1}{2} \mu_\gamma ^2 A_\mu A^\mu
\end{equation}
where $\mu_\gamma ^{-1}$ is the reduced Compton wavelength associated
with the photon rest mass and $j_\mu=\left(\rho,-\vec{j}\right)$ is
the 4-current. $F^{\mu\nu}$ is the electromagnetic tensor and $A_\mu$
the 4-vector potential. From the Lagrangian density, one obtains the
covariant form of the Proca-Maxwell equations
\begin{equation}
\frac{\partial F_{\mu\nu}}{\partial x_\nu}+\mu_\gamma ^2 A_\mu=4\pi j_\mu
\label{eq3}
\end{equation}
that leads to the Proca wave equation for $A_\mu$
\begin{equation}
(\dalembert -\mu_\gamma ^2)A_\mu= - 4 \pi j_\mu
\end{equation}

When one considers a photon propagating in a plasma, the usual Maxwell
equations can be replaced by the set of Proca-Maxwell equations
in which there appears a mass-like term for the photon due to
light-matter interaction \cite{hora1991plasmas}. The usual formulation
of Proca-Maxwell equations is obtained by expanding Eq. (\ref{eq3}) in
terms of the electric $\vec{E}$ and magnetic $\vec{B}$ fields, that, in
the presence of charges and currents $\rho$ and $\vec{j}$, are
\begin{align}
 &\div\vec{E} = 4\pi\rho-\mu_\gamma^2\phi
\\
 &\curl\vec{E} = -\frac{\partial\vec{B}}{\partial t}\nonumber
\\
 &\div\vec{B} = 0 \nonumber
\\
 &\curl\vec{B} = 4\pi\vec{j} + \frac{\partial\vec{E}}{\partial t}
  - \mu_\gamma^2\vec{A} \nonumber
\end{align}
and for $\mu_\gamma\rightarrow 0$, they smoothly reduce to Maxwell's equations. 
The Poynting vector for massive photons depends directly on both the scalar and vector potentials
\begin{equation}
\vec{S}=\frac 1{4 \pi}(\vec{E}\cross\vec{B}
 + \mu_\gamma^2\phi\vec{A})
\end{equation}
and also the energy density has an explicit dependency on the potentials
\begin{equation}
u=\frac{1}{8 \pi}\left(\vec{E}^2 + \vec{B}^2 + \mu_\gamma^2\phi^2
 + \mu_\gamma^2\vec{A}^2\right)
\end{equation}
While the Lorentz invariance remains valid, the gauge invariance is
lost because the potentials become observable through the
acquired energy densities $\mu_\gamma^2\phi^2/8\pi$ and $\mu_\gamma ^2
\vec{A}^2 /8\pi$, respectively, due to the interaction of the photon with the plasma or
its confinement within the observable Universe.

Let us consider a transverse EM wave propagating through an isotropic
plasma, cast to form a helicoidal static plasma vortex. The heavy ions
constitute a neutralising background, and their motion can, in the
first approximation, be neglected. The motion of free electrons forms
a three-dimensional current $\vec{j} = - n\vec{v}$, where 
$n$ is the electron number density, and $\vec{v}$ the
velocity of the electrons in the medium derived from the electron fluid
equations
\begin{align}
&\frac{\partial n}{\partial t}+ \div n\vec v=0
\\
&\frac{\partial \vec v}{\partial t}+ \vec v \bdot \grad \vec v
 = - \frac{1}{m} (\vec E + \vec v\cross\vec B)
\end{align}
Thermal and relativistic mass effects are ignored. In the first
approximation, the mean electron velocity in the static plasma vortex
becomes $\vec v=\vec v_0(\vec r,t)+\delta \vec v$ where $\vec v_0$
is the background velocity and $\delta \vec v$ is the perturbation
associated with the propagating EM wave. The electron number density is
\begin{equation}
n=n_0 + \tilde{n}(r,z)\cos(l_0\varphi+q_0 z)
\end{equation}
where $n_0$ is the background plasma density and the helicoidal density
perturbation in the plasma is given by the latter term, expressed in
cylindrical coordinates, $\vec r\equiv(r,\varphi,z)$. The coordinate
$z$ is the axis of symmetry around which the electron spiral is winding
and along which the EM wave is propagating. 

Similarly as in a spiral phase plate, the number of electrons affecting the EM wave
depends on $z$ and can vary slowly on a scale much larger than the
spatial period $z_0=2\pi/q_0$, where $q_0$ is the helix step. For a
typical double vortex one obtains $l_0=1$. Neglecting for a moment
the rotation of plasma and considering the case of a static helical
perturbation, the current density perturbation in the plasma is
$\vec{j}=- n(\vec{r})\delta\vec{v}$, and the propagation equation of
the electric field becomes
\begin{equation}
\label{ewave}
\bigg(\frac{\partial^2}{\partial t^2} - \nabla^2 - \omega_{p}^2\bigg)\vec E=0
\end{equation}
where $\omega_p^2= \omega_{p0} ^2[1+\epsilon (r,\varphi,z)]$
and $\omega_{p0}$ is the unperturbed plasma frequency, given by
$\omega_{p0}^2=4\pi n_0/m$. The quantity $\epsilon$ represents the
perturbation. We consider solutions of the form
\begin{equation}
\vec E(\vec r,t)=\vec a(\vec r)
 \exp\Big(-\im\omega t+\im\int^z k(z')\,\d{z'}\Big)
\end{equation}
where $\omega$ is the EM wave frequency and $\vec a(\vec r)$ is
the amplitude, varying slowly along $z$ such that
\cite{Mendonca&al:PRL:2009}
\begin{equation}
\left|\frac{\partial^2 \vec a}{\partial z ^2}\right| \ll \left|2k\frac{\partial \vec a}{\partial z}\right|
\end{equation}
We can then write the wave equation in a perturbed paraxial equation form
\begin{equation}
\Big(\grad_\bot ^2 + 2ik\frac{\partial}{\partial z}-\omega_{p}^2\Big)\vec a=0
\end{equation}
and the dispersion relation, connecting $k$ and $\omega$, is
\begin{equation}
k^2=\omega^2-\omega_{p0}^2 \, \left[1+\epsilon (r,\varphi,z) \right] \, .
\end{equation}

In our case, a general solution to the wave equation in the
paraxial approximation can be represented in a basis of orthogonal
L-G modes, having the amplitude
\begin{equation}
\vec a(r,\varphi,z)=\sum_{pl} b_{pl}(r,z)e^{\im l\varphi}
 \exp\Big(-\frac{r^2}{2w^2}\Big)\hat{\vec{e}}_{pl}
\end{equation}
where $w\equiv w(z)$ is the beam waist, $\hat{\vec{e}}_{pl}$ are unit
polarisation vectors, $A_{pl}$ are the amplitudes of each mode and
\begin{equation}
b_{pl}(r,z)
 =c_{pl}(z)\sqrt{\frac{(l+p)!}{4\pi p!}}
 \Big(\frac{r^2}{w^2}\Big)^{\!|l|}\,
 L_p^{|l|}\Big(\frac{r^2}{w^2}\Big)
\end{equation}
where the function $L_p ^{| l |}$ represents the associated Laguerre polynomial. As usual, the integers $p$ and $l$ are the radial and the azimuthal quantum numbers, respectively. The electric field then becomes
\begin{equation}
\vec E(\vec r,t)=\sum_{pl} \vec E_{pl}(\vec r)
 \exp\Big(-\im\omega t + \im\int^z k(z')\,\d{z'}\Big)
\end{equation}
with
\begin{equation}
\vec E_{pl}(\vec r)=c_{pl}(z) F_{pl}(r,\varphi)\hat{\vec{e}}_{pl}
\end{equation}
where $F_{pl}(r,\varphi)$ is given by formula~(\ref{eqn:lgmode}).

When a vortex perturbation $\varepsilon (r,\varphi, z)$ is present, the
modes will couple according to
\begin{equation}
\frac{\partial}{\partial z}c_{pl}(z)
 =\frac{\im}{2k}\sum_{p'l'}K(pl,p'l')c_{p'l'}
\end{equation}
where $K(pl,p'l')$ are the coupling coefficients. If the EM wave does
not carry OAM and the mode coupling is sufficiently weak that the
zero OAM mode dominates over the entire interaction region, one obtains a
coupling
\begin{equation}
K(pl,p'l')=\pi \omega_{p0} ^2 \frac{\tilde{n}}{n_0}\delta_{pp'}[\delta_{l',-l_0}e^{\im q_0z}+\delta_{l',l_0}e^{-\im q_0z}]
\end{equation}
If we assume the same polarisation state for all the interacting modes,
the field mode amplitudes describe the rate of transfer of OAM from the
static plasma vortex to the EM field
\begin{equation}
c_{p,\pm l_0}(z)
 =\im \frac{\pi c(0)}{2c^2} \int_0 ^z \frac{\omega_{p0}^2(z')}{k(z')}
 \frac{\tilde{n}(z')}{n_0}e^{\mp \im q_0z'}\,\d{z'}\,.
\end{equation}
A more general solution, where the amplitude of the
initially excited mode is allowed to change, is discussed in
Ref.~\cite{Mendonca&al:PRL:2008}. The initial OAM state $l_i$ of
the electromagnetic beam decays into other states $(l_i+ul_0)$ on a
length scale approximately determined by the inverse of the coupling
constant, showing an effective exchange of OAM states between the photons
and the plasma.

The photon mass in Proca equations is defined as $m_\gamma=\mu_\gamma$
and the effective photon mass is also related to the plasma frequency
$m_{\text{eff}}=\omega_p$ \cite{PhysRev.130.439}. By comparing these two definitions, one obtains
the equivalence between the inverse of the characteristic length in a
plasma, $\mu_\gamma$ and the plasma frequency, $\mu_\gamma=\omega_p$.
When considering the EM wave equation of the Proca field one obtains a Klein-Gordon equation for the 4-vector potential 
\begin{equation}
(\dalembert -\mu_\gamma ^2)A_\mu=-4 \pi j_\mu
\end{equation}
with the constraint derived from the massive photon in a medium
($\partial^\mu{A}_\mu=0$). By differentiating this expression with
respect to time, and considering, in addition to the previous
calculations, the simplest case where $\mu_\gamma$ is a constant in time, one
obtains
\begin{equation}
(\dalembert -\mu_\gamma ^2) \frac \partial{\partial t}A_\mu=-4\pi  \frac \partial{\partial t} j_\mu
\end{equation}

We now consider the spatial components applied to the case of photons
moving in a plasma. In particular, we consider a plasma with a
well-defined structure of a static plasma vortex, in which the density
$n=n(\vec r)$ is not a function of time. Being the set of equations
independent on the temporal part, we pass from the four-dimensional
notation to a three-dimensional vectorial notation without losing in
generality. The current density has the form $\vec j= -n(\vec r) \vec
v$, and its derivative becomes
\begin{equation}
\frac \partial{\partial t} j_i
 =-\Big(\delta\vec v \frac \partial{\partial t} n(\vec r)
 +n(\vec r)\frac \partial{\partial t} \delta \vec v\Big)
=-n(\vec r) \frac \partial{\partial t} \vec v
\end{equation}
and the wave equation of the Proca EM field becomes
\begin{equation}
(\dalembert -\mu_\gamma ^2) \frac \partial{\partial t} \vec A=-4 \pi n(\vec r)\frac \partial{\partial t}\vec v.
\end{equation}
Using $\vec E=-\grad\phi-\frac \partial{\partial t}\vec A$ one
obtains the wave equation for the EM field
\begin{equation}
(\dalembert -\mu_\gamma ^2) (\vec E+\grad\phi)= 4\pi n(\vec r)\frac \partial{\partial t}\vec v.
\end{equation}
Assuming that $\vec v$ is parallel to $\vec E$, and $\vec v=\vec
v_0+\delta\vec v$, where $\delta \vec v$ is the perturbation of
electrons velocity associated with the propagating EM wave, we obtain
for $\vec E\neq0$ the wave equation for the EM field in the plasma,
\begin{equation}
\bigg[\dalembert
 - \mu_\gamma^2\bigg(1+\frac{\hat v\bdot\grad\phi}{|\vec E|}\bigg)
 - 4 \pi \frac{n(\vec r)\dot{\delta v}-\hat{v}\bdot\dalembert\grad\phi}
 {|\vec E|}\bigg]\vec E=0
\end{equation}
where $\hat{v}= \vec v /|v|$ is the direction vector of the velocity field, $\hat v$ the unit vector of velocity and $\dot{\delta v}=\hat v\bdot\partial_t v=|\partial_t v|$.
When comparing this equation with Eq .~(\ref{ewave}), derived from the
electric field propagation equation in the case of a static plasma
vortex, one obtains
\begin{equation}
\mu_\gamma^2\left(1+\frac{\hat v\bdot\grad\phi}{E}\right)
 + 4\pi\frac{n(\vec r)\dot{\delta v}-\hat{v}\bdot\dalembert\grad\phi}{E}
 =\omega_{p} ^2
\end{equation}
that implies a direct relationship between the effective mass that
a photon acquires in a plasma, the plasma frequency and the orbital
angular momentum because of the peculiar spatial distribution:
\begin{align}
\begin{split}
\mu_\gamma^2
={}&
 \frac{E}{E+\hat{v}\bdot\grad\phi}\omega^2_{p0}[1+\varepsilon(r,\varphi,z)]
\\
  &-\frac{1}{E+\hat{v}\bdot\grad\phi}\Big(
  4\pi\dot{\delta v}\big[n_0+\tilde{n}\cos(l_0\varphi+q_0z)\big]
\\
  &\qquad\qquad\qquad- 4 \pi \hat{v}\bdot\dalembert\grad\phi\Big)
\end{split}
\end{align}

The new mass component is a fictitious term, generated by the
interaction of photons with the plasma and cannot be ascribed to an
intrinsic property of the photon.

When the electron number density exhibits certain spatial properties,
such as vortices, any photon has an associated virtual mass term that is
smaller than that expected from Proca equations in a homogeneous plasma,
because of a negative term that corresponds to a precise orbital angular
momentum component.

\section{Conclusions}

We have investigated the problem of photon mass in a plasma and show
that part of the acquired mass term is related to the orbital angular
momentum of light imposed by certain spatial distributions of the
plasma. We focused our attention on the simplest case of spatial
distributions described by a static plasma vortex. This approach shows
that also the spatial distribution of charges can impose OAM and an
additional mass term that reduces the effective mass of the photon
inside a non-structured plasma, degrading the variation in OAM. The
spiral-like plasma structure induces OAM states in most of the scattered
photons and the stochastic interaction value responsible for the Proca
mass term is instead transformed into an organised state of light, with
the result of reducing the averaged mass term \cite{bookOAM2}.

\acknowledgments
One of the authors (B.\,T.), gratefully acknowledges the financial
support from the Swedish Research Council (VR). F.T. acknowledges the financial support of CARIPARO in the 2006 program of excellence.


\end{document}